\documentclass[11pt]{article}

\usepackage{epsfig,amssymb,amstext}     

\newcommand{\nhim}{{S_{\text{NHIM}}^{2n-3}(E)}}
\newcommand{\ts}{{S_{\text{ts}}^{2n-2}(E)}}
\newcommand{\tsdef}{{\tilde{S}_{\text{ts}}^{2n-2}(E)}}
\newcommand{\tsf}{{D_{\text{ts,\,f}}^{2n-2}(E)}}
\newcommand{\tsb}{{D_{\text{ts,\,b}}^{2n-2}(E)}}
\newcommand{\po}{{S_{\text{NHIM}}^1(E)}}
\newcommand{\podef}{{\tilde{S}^1(E)}}
\newcommand{\tstwoDOF}{{S_{\text{ts}}^{2}(E)}}
\newcommand{\tsftwoDOF}{{D_{\text{ts,\,f}}^{2}(E)}}

\newcommand{\tstwoDOFdef}{{\tilde{S}_{\text{ts}}^{2}(E)}}
\newcommand{\tsftwoDOFdef}{{\tilde{D}_{\text{ts,\,f}}^{2}(E)}}
\newcommand{\tsbtwoDOFdef}{{\tilde{D}_{\text{ts,\,b}}^{2}(E)}}

\newcommand{\R}{{\mathbb R}}

\newcommand{\rem}[1]{}

\begin{document}

\thispagestyle{plain}
\noindent
{\Large Direct Construction of a  Dividing Surface of Minimal Flux for Multi-Degree-of-Freedom Systems: The Equivalence of Conventional and Variational Transition State Theory} \\[5ex]
\begin{center}
{\large H. Waalkens and  S. Wiggins}\\[4
ex]
{\large School of Mathematics\\
        Bristol University\\ University Walk\\
        Bristol, BS8 1TW , UK}\\[2ex]

\today
\end{center}

\vspace*{1ex}

\section*{Abstract}
The fundamental assumption of conventional transition state theory is the existence of a dividing surface having the property that trajectories originating in reactants must cross the surface {\em only once} and then proceed to products. Recently it has been shown \cite{wwju, ujpyw} how to construct a dividing surface in {\em phase space} for Hamiltonian systems with an arbitrary (but finite) number of degrees of freedom having the property that trajectories only cross  once {\em locally}. In this letter we provide an argument showing that the flux across this dividing surface is a minimum with respect to certain types of variations of the dividing surface. In this sense, conventional transition state theory is  shown to be equivalent to variational transition state theory.

\vspace{3ex}

\noindent PACS: 82.20.Db,  82.20.Nk,  05.45.-\\


\section{Introduction}

Transition state theory provides a fundamental framework for computing chemical reaction rates.  The original ideas are due to Wigner, Polanyi, and Eyring, yet much work on different aspects of the subject continue to this day (see the recent reviews of \cite{miller} and \cite{truhlar1}).  In recent years transition state theory has been shown to be much more widely applicable than just for problems in chemical reactions. It has been used in atomic physics  \cite{JaffeFarellyUzer2000}, studies of the rearrangements of clusters \cite{KomaBerry1999}, solid state and semi-conductor physics \cite{Jacucci,Eckhardt1995}, cosmology \cite{cosmo}, and  celestial mechanics \cite{ross}.

The fundamental assumption of transition state theory is the existence of a dividing surface having the property that trajectories originating in reactants must cross the surface {\em only once} and then proceed to products.
The reaction rate is proportional to the flux through the dividing surface. The construction of such a surface in specific systems, especially those with more than two degrees of freedom (DOF),  has posed extreme difficulties. The phenomenon of recrossing of a dividing surface gives rise to a larger value for the flux, which naturally inspires the idea of varying the dividing surface so that the flux is minimised, this has lead to the subject of variational transition state theory
(\cite{Keck1967}, see also the review of \cite{truhlar2}).

\begin{figure}
\centerline{
\raisebox{4cm}{(a)}
\includegraphics[angle=0,height=4cm]{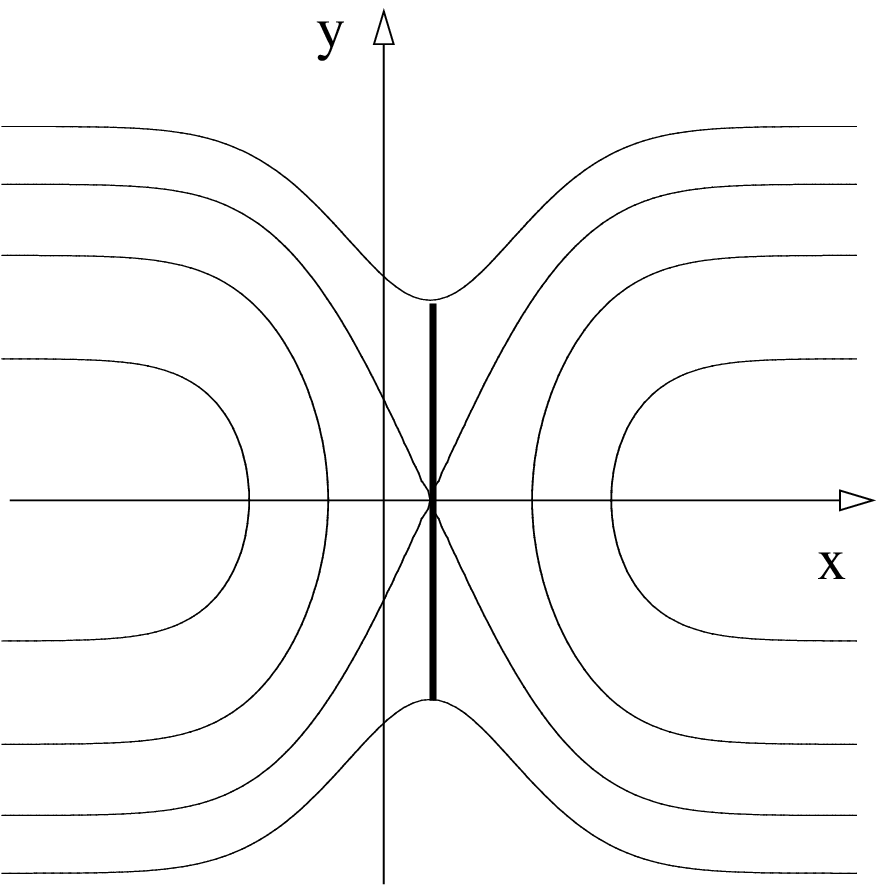}
\hspace*{1cm}
\raisebox{4cm}{(b)}
\includegraphics[angle=0,height=4cm]{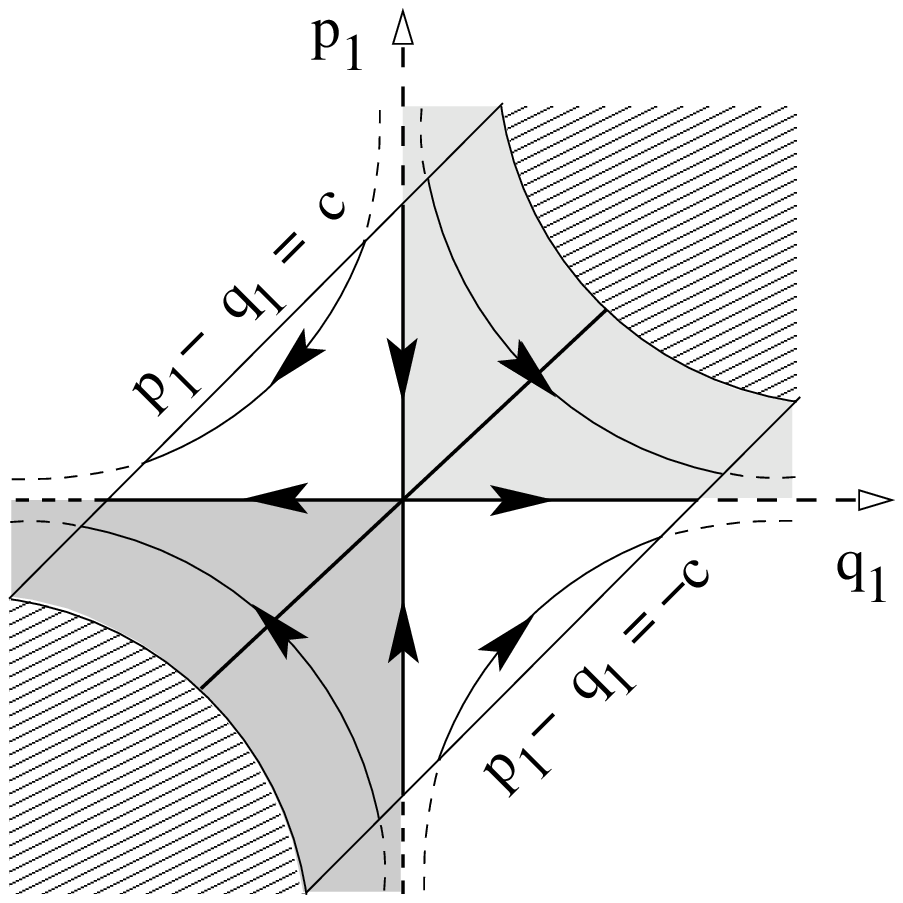}
}
\caption{\label{fig:pods}
(a) Equipotentials and PODS (bold line) near a saddle. 
(b) Saddle plane $(q_1, p_1)$ with projections of the $(2n-1)$-dimensional energy surface $H=E$ for $E>0$ (all but the hatched
region), the stable and unstable spherical cylinder (the $p_1$-axis and the $q_1$-axis), 
the NHIM (the origin), and the $(2n-2)$-spheres 
$p_1-q_1=0$ (dividing surface). 
The light and dark grey regions are the
projections of the energy surface volumes enclosed by the forward and backward reactive spherical cylinder 
$W_f(E)$ and $W_b(E)$, respectively. $p_1-q_1=\pm c$, $c>0$, delimit the region of validity of the normal form.
}
\end{figure}

For systems with 2  DOF described by a Hamiltonian $H$ of the simple type kinetic-plus-potential
the problem of constructing a dividing surface (transition state) with trajectories only crossing once and having minimum flux was solved during the 70's by 
McLafferty, Pechukas and Pollak \cite{pm, pp1, pp2, pp3}. They considered the line in configuration space given by the projection of an unstable periodic orbit (the Lyapunov periodic orbit associated with a saddle point), see figure~\ref{fig:pods}(a). 
This line which is referred to as {\sl periodic orbit dividing surface} (PODS) separates the reactant region ($x<0$) from the product region $(x>0)$ by  connecting two pieces of an equipotential. This solves the problem of (local) recrossings as there is no trajectory evolving from reactants to products (or vice versa) whose projection to configuration space is tangent to this line. After  a  reacting trajectory has crossed the PODS it has to leave the neighbourhood of the PODS before it can possibly recross it.

To comprehend a PODS as a {\em phase space} object 
we rewrite the
energy equation as
\begin{equation}
\label{eq:pods}
p_x^2 + p_y^2 = 2m(E-V(x,y))\,.
\end{equation}
For $(x,y)$ on the projection of the periodic orbit in figure~\ref{fig:pods}(a) the equation in (\ref{eq:pods}) defines a family of circles 
in $(p_x,p_y)$ which shrink to points at the turning points of the periodic orbit, i.e. the PODS is a 2-sphere.
The periodic orbit which has $p_x=0$  can be considered as the equator of the PODS. It divides the PODS into two hemispheres
which have $p_x>0$ and $p_x<0$, respectively. 
The flux from $x<0$ (``reactants'') to $x>0$ (``products'') through the PODS in figure~\ref{fig:pods}(a) is commonly written as
\begin{equation}
\label{eq:flux_pods}
{\cal F}_E = \int \mbox{d} x \,\mbox{d} y \, \mbox{d} p_x \, \mbox{d} p_y  \, \delta(E-H) 
\delta(x-x_0) \Theta(p_x) \frac{p_x}{m}  \,.
\end{equation}
The step function $\Theta(p_x)$ effectively restricts the integral to one hemisphere of the PODS.
Using Stokes' theorem it is not difficult to see that 
the flux in (\ref{eq:flux_pods}) is given by the action of the periodic orbit (p.o.), i.e.
\[
{\cal F}_E =
\oint_{p.o.} {\bf p} \,\mbox{d}{\bf q} \,.
\]

Unfortunately, the methods of McLafferty, Pechukas and Pollak are specific to 2 DOF, and as well to Hamiltonian's of a certain form.
Recently it has been shown \cite{wwju, ujpyw} how to construct a dividing surface in {\em phase space} for Hamiltonian systems with an arbitrary (but finite) number of DOF having the property that trajectories only cross  once locally. A computational algorithm for realising this construction was also given. However, it was not shown that this transition state was  a surface of minimum flux. In this letter we give a geometrical argument demonstrating that this is indeed the case, and in the process show that for the phase space transition states constructed according to the techniques in  \cite{wwju, ujpyw}, conventional transition state theory is equivalent to variational transition state theory.  In order to do this we must first recall the elements of the many DOF theory in  \cite{wwju, ujpyw}.


\section{Dividing surface for  multi-dimensional systems}

In the general case of $n$ DOF we consider an equilibrium point of saddle-center-...-center
type, i.e. the linearised vector field about the equilibrium point has one pair of real eigenvalues $\pm \lambda$ and $n-1$ imaginary eigenvalues $\pm i\omega_k$, $k=2,\dots,n$,  where without restriction of generality $\lambda,\omega_k>0$.
Equilibria of this type are characteristic for systems with a reaction type dynamics, and they occur in all of the examples given in the introduction.
The transport is controlled by various high-dimensional manifolds which can be realised and computed through a procedure based on Poincar{\'e}-Birkhoff normalisation. In fact, under general conditions (see \cite{ujpyw} for the details), near a saddle-center-...-center one 
can construct a sequence of local, non-linear, symplectic transformations of the phase space coordinates that transform the Hamiltonian into

\begin{eqnarray}
\label{eq:normalform}
H = \lambda p_1 q_1 +
\sum_{k=2}^n \frac{\omega_k}{2}(p_k^2+q_k^2) 
& + & f_1({\cal I},q_2,\dots,p_n,q_2,\dots,q_n) \nonumber \\  
& + & f_2(q_2,\dots,p_n,q_2,\dots,q_n)
\end{eqnarray}

\noindent
up to any desired order. Here $(q_1,\dots,q_n,p_1,\dots,p_n)$ are canonical phase space coordinates, ${\cal I}=p_1\,q_1$ and $f_1$, $f_2$ are at least of third order, i.e. they are responsible for the nonlinear term in the Hamiltonian vector field.  Moreover $f_1$ has the property that it vanishes for ${\cal I}=0$.
$(q_1,p_1)$ play the role of reaction coordinates. $(q_2,\dots,q_n,p_2,\dots,p_n)$ are the bath coordinates.

The importance of saddle-center-...-center equilibria 
for reaction type dynamics can be inferred from the topology of energy surfaces $H=E$. First consider the quadratic Hamiltonian 
given by the first part in (\ref{eq:normalform})
and  write the energy equation as

\begin{equation}
\label{eq:top_en}
E +  \frac{\lambda}{4} (p_1-q_1)^2 = 
\frac{\lambda}{4} (p_1+q_1)^2 + 
\sum_{k=2}^{n} \frac{\omega_k}{2} (p_k^2 + q_k^2)\,.
\end{equation}

\noindent
For $E<0$ the left hand side is positive for $p_1-q_1<-(-4E/\lambda)^{1/2}$ or $p_1-q_1>(-4E/\lambda)^{1/2}$. 
For a fixed $p_1-q_1$ in either of these ranges 
the equation in (\ref{eq:top_en}) defines a $(2n-2)$ dimensional sphere $S^{2n-2}$.
For $p_1-q_1=\pm (-4E/\lambda)^{1/2}$ the $(2n-2)$-spheres shrink to points. 
The energy surface thus appears as two disjoint {\em spherical cones} which correspond to ``reactants'' and ``products'',
respectively.
Increasing the energy to 
$E>0$, the left hand side of (\ref{eq:top_en}) is strictly positive. The 
formerly disjoint components merge and the energy surface becomes a {\sl spherical cylinder}  $S^{2n-2} \times \R$.
Restricting to a sufficiently small neighbourhood 
by confining $p_1-q_1$ to an interval $I=[-c,c]$ with $c>0$ 
sufficiently small 
and with $E$ sufficiently close to zero, 
the topological consideration remain true if the non-linear terms $f_1$ and $f_2$ are taken into account. 
Moreover, for a high but finite order normal form, the error from neglecting the non-normalised ``tail'' of the Hamiltonian can be made as small as desired by choosing the interval $[-c,c]$ sufficiently small and $E$ sufficiently close to zero.

The importance  of the normal form arises from the fact that it gives explicit expressions for all the manifolds which control the dynamics from
reactants to products. For a fixed energy $E>0$ these manifolds are

\begin{itemize} 
\item {\em The saddle sphere $\nhim$}:
On the energy surface $H=E$ the equation $p_1=q_1=0$ defines a $(2n-3)$-sphere which we denote by 
$\nhim$. 
It can be considered as a ``big saddle''. In fact, it
is a so-called {\em normally hyperbolic invariant manifold} (NHIM) where 
normal hyperbolicity means that the expansion and contraction rates transverse to the manifold 
dominate those tangent to the manifold.

\item {\em The forward and backward reactive spherical cylinders $W_f(E)$ and $W_b(E)$}:
The saddle sphere has stable and unstable manifolds $W^s(E)$ and $W^u(E)$ which are iso-energetic, i.e. contained in the energy surface, 
and which are explicitly 
given by $q_1=0$ and $p_1=0$, respectively. 
These invariant manifolds have the topology of spherical cylinders 
$S^{2n-3}\times I$. Since they  are of codimension 1 in the energy surface, i.e. they are of one dimension less than the energy surface, 
they act as impenetrable barriers. 
$W^s(E)$ and $W^u(E)$ each appear as two branches. We call the branch of  $W^s(E)$ which has $p_1>0$ 
the {\em forward branch} $W^s_f(E)$ and the
branch which has $p_1<0$ the {\em backward branch} $W^s_b(E)$. 
Likewise the forward branch $W^u_f(E)$ of $W^u(E)$ has $q_1>0$ and the backward branch $W^u_b(E)$ has $q_1<0$.
We call the union  of the forward branches, $W_f(E) := W^s_f(E)\cup W^u_f(E)$, the {\em forward reactive spherical cylinder}. Similarly, 
we call the union of the backward branches, $W_b(E):=W^s_b(E)\cup W^u_b(E)$,  the {\em backward reactive spherical cylinder}. 
The significance of these spherical cylinders arises from the fact that they enclose  volumes of the energy surface which contain all forward and 
all backward reactive trajectories, respectively. All non-reactive trajectories 
are contained in the complement of these two volumes.
\end{itemize}
 
We define the dividing surface or transition state as follows.
\begin{itemize} 
\item {\em Transition state $\ts$}: 
On the energy surface $H=E$ the equation $p_1=q_1$ defines a $(2n-2)$-sphere
which we denote by $\ts$. It is of codimension 1 in the energy surface. It divides the energy surface into two
components: the reactant region  $p_1-q_1>0$ and the product region $p_1-q_1<0$. The saddle sphere $\nhim$ can be considered as the ``equator'' 
of the transition state.
It divides $\ts$ into the two hemispheres: the  {\em forward hemisphere} $\tsf$, which has $p_1=q_1>0$, and the {\em backward hemisphere} $\tsb$, which has $p_1=q_1<0$. 
$\tsf$ and $\tsb$ are topological $(2n-2)$-discs.
Except for its equator, which is an invariant manifold, the transition state is everywhere transverse to the Hamiltonian flow
as is easily seen from the equations of motions 
derived from the normal form 
Hamiltonian (\ref{eq:normalform}). This means that a trajectory, after having crossed the transition state, has to leave the neighbourhood of the 
transition state before it can
possibly cross it again, i.e.  the transition state {\em locally} is a  ``surface of no return''.
\end{itemize}

An important advantage of the normal form coordinates is that dynamical issues related to flux and reactivity  can be understood to a great extent from the projection to 
the plane of the reaction coordinates $(q_1,p_1)$, see figure~\ref{fig:pods}(b). 
Due to the constance of the saddle integral ${\cal I}=p_1 q_1$ trajectories project to hyperbolas.
Forward and backward reactive trajectories project to the first quadrant $q_1,p_1>0$. 
It approaches the forward hemisphere $\tsf$ of the transition state inside of the forward branch of the stable cylinder 
$W^s_f(E)$ whose interior projects to the part of the first quadrant above the diagonal $p_1=q_1>0$. After crossing the transition state the trajectory leaves inside of the forward branch of the unstable spherical cylinder $W^u_f(E)$ whose interior projects to the part of the first quadrant below the diagonal $p_1=q_1>0$. While similar considerations hold for the backward reactive trajectories, non-reactive trajectories project to the second quadrant $p_1>0$, $q_1<0$ which corresponds to reactants or to the fourth quadrant $p_1<0$, $q_1>0$ which corresponds to products.  Therefore reactive trajectories have 
${\cal I}=p_1q_1>0$ and  non-reactive trajectories have ${\cal I}=p_1q_1<0$.


\section{Minimal flux property of the transition state}

Consider at first the phase space volume form $\Omega=\mbox{d}p_1\wedge\mbox{d}q_1\wedge\cdots\wedge\mbox{d}p_n\wedge\mbox{d}q_n$, 
which in terms of the symplectic 2-from  $\omega=\sum_{k=1}^n \mbox{d} p_k \wedge \mbox{d} q_k$ can be written as $\Omega=\omega^n/n!$. 
Let $\eta$ be an energy surface volume form defined via the property $\mbox{d}H\wedge \eta = \Omega$.
Then the flux through a codimension 1 submanifold of the $(2n-1)$-dimensional energy surface $H=E$
is obtained  from integrating over it the ``flux'' form $\Omega'$ given by the interior product of the Hamiltonian vector field $X_H$ with $\eta$
\cite{MacKay1990}, i.e.

\begin{equation}
\label{eq:omegaprime}
\Omega' =i_{X_H}\eta = \frac{1}{(n-1)!} \omega^{n-1}
\end{equation}
where $i_{X_H}\eta(\xi_1,\dots,\xi_{2n-2})=\eta(\xi_1,\dots,\xi_{2n-2},X_H)$ for any $2n-2$ vectors $\xi_k$. The second equality 
in (\ref{eq:omegaprime}) is easily established on a non-critical energy surface, i.e. on an energy surface 
which contains no equilibria.
The flux form $\Omega'$ is exact. In fact the generalised ``action'' form

\[
\varphi = \sum_{k=1}^n p_k\mbox{d} q_k \wedge \frac{1}{(n-1)!} \omega^{n-2}
\]

\noindent
has the property $\mbox{d} \varphi = \Omega'$ and facilitates the use of Stokes' theorem 
to compute the flux.
In the case of two DOF we simply have
$
\Omega' = \omega=\mbox{d}p_1 \wedge \mbox{d}q_1 + \mbox{d}p_2 \wedge \mbox{d}q_2
$
and
$
\varphi = p_1 \mbox{d}q_1 + p_2 \mbox{d}q_2\,.
$
Since the transition state $\ts$ is a sphere, that is, a manifold without boundary, it follows from Stokes' theorem 
that the integral of $\Omega'$ over  $\ts$ is zero. 
As in the case of PODS one has to distinguish between the directions in
which the Hamiltonian flow crosses the transition state. 
Given a  normal bundle over $\ts$
the direction can be specified by the sign of the scalar product
between the normal vectors and the Hamiltonian vector field. 
This scalar product is strictly positive on one hemispheres of $\ts$, 
strictly negative on the other hemisphere and zero only at the equator of $\ts$, i.e. at the saddle sphere $\nhim$, 
where the Hamiltonian vector field is
tangent to $\ts$.  
Likewise, the flux form $\Omega'$ on $\ts$ vanishes nowhere on $\tsf$ and $\tsb$ and is identically zero on $\nhim$.
It is natural to take as the orientation of $\tsf$ and $\tsb$ the orientation they inherit from the
transition state. Without restriction we may assume that the orientation of $\ts$ is such that $\Omega'$ is positive on the forward hemisphere $\tsf$ and
negative on the backward hemisphere $\tsb$, i.e. $\Omega'$ and $-\Omega'$ can be considered as {\em volume forms} on $\tsf$ and $\tsb$, 
respectively.
It follows from Stokes' theorem that
the flux through the forward and backward hemispheres, $\int_\tsf \Omega'$ and $\int_\tsb \Omega'$, have the same magnitude but opposite sign and
can be computed from integrating the action form $\varphi$ over the saddle sphere: $\int_\tsf \Omega' =-\int_\tsb \Omega' = |\int_\nhim \varphi|$. We
call the positive quantity $\int_\tsf \Omega'$ the {\em forward flux} and 
the negative quantity $\int_\tsf \Omega'$ the {\em backward flux} through $\ts$ .

We now show that the forward flux through $\ts$ is minimal. This  can be stated as a variational problem
and a result in this direction is obtained by MacKay \cite{MacKay1991} who proves that the integral $\int_C \varphi$ of the action 
form $\varphi$ over codimension 2 submanifolds $C$ of the energy surface is stationary
with respect to variations of $C$ 
if and only if $C$ is invariant under the Hamiltonian flow. Since the saddle sphere $\nhim$ is an invariant manifold, MacKay's result implies 
that the ``action'' of $\nhim$, $\int_\nhim \varphi$, is stationary  with respect to variations of $\nhim$. 
MacKay considers arbitrary variations within the energy surface and this leads to an indefinite variational principle.
In fact, in the context of transition state theory it is more useful to consider 
variations of the codimension 1 transition state $\ts$, 
which in a sense that will become clear below, imply variations of the codimension 2 saddle sphere $\nhim$ in a smaller, 
more suitable class than in MacKay's case.

\begin{figure}
\centerline{
\raisebox{3.7cm}{(a)}
\includegraphics[width=4cm]{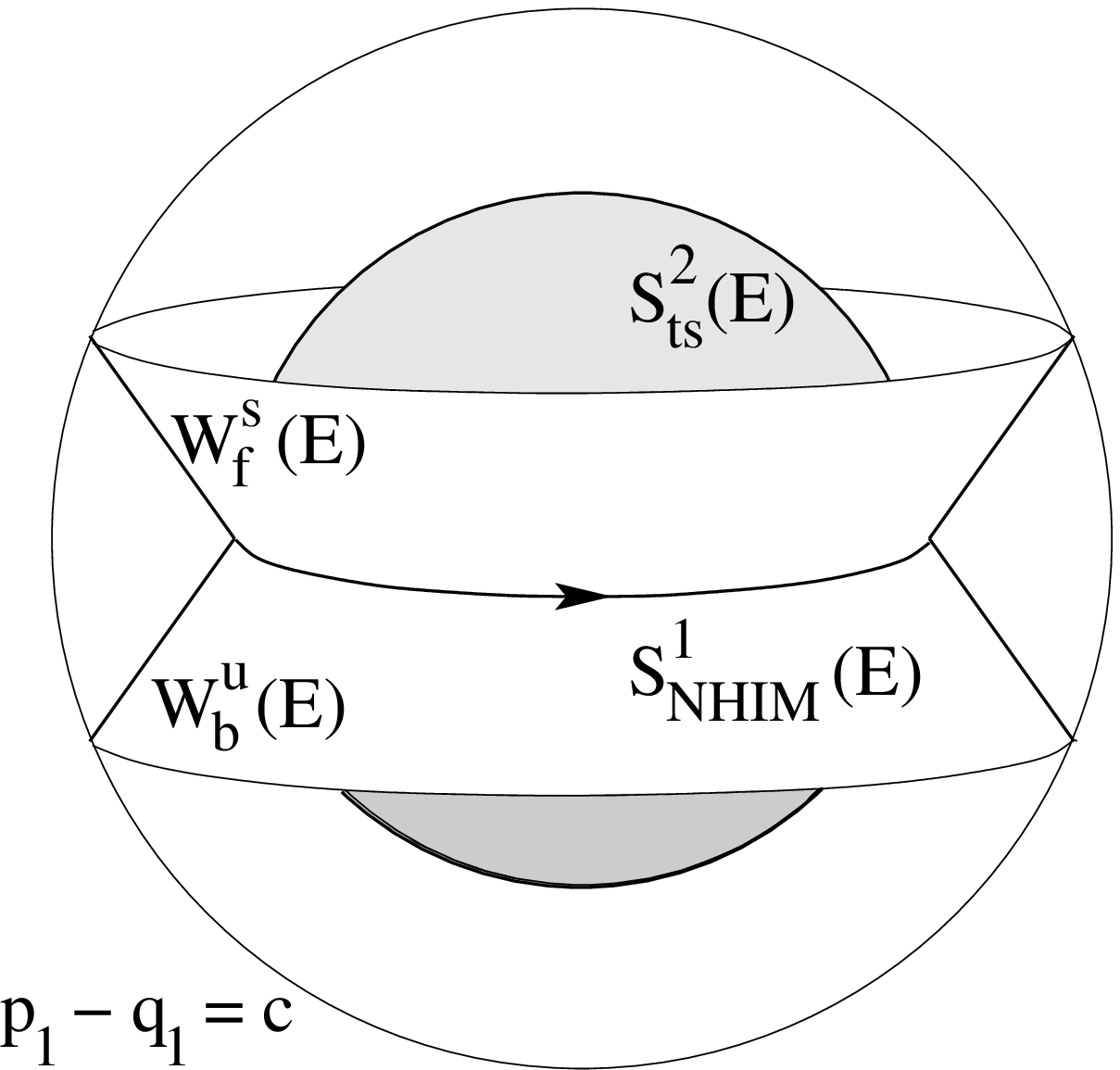} 
\raisebox{3.7cm}{(b)}
\raisebox{0.7cm}{\includegraphics[width=4cm]{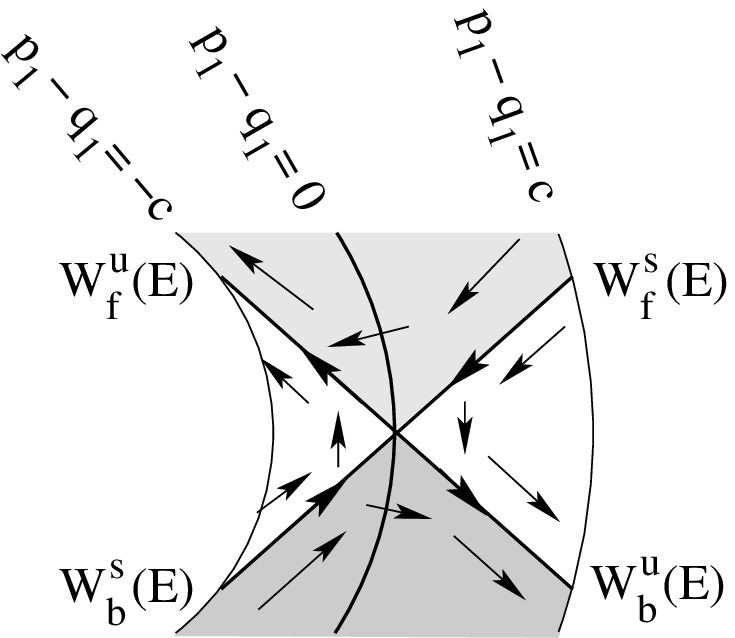}} 
}
\vspace*{0.3cm}
\centerline{
\raisebox{3.7cm}{(c)}
\includegraphics[width=4cm]{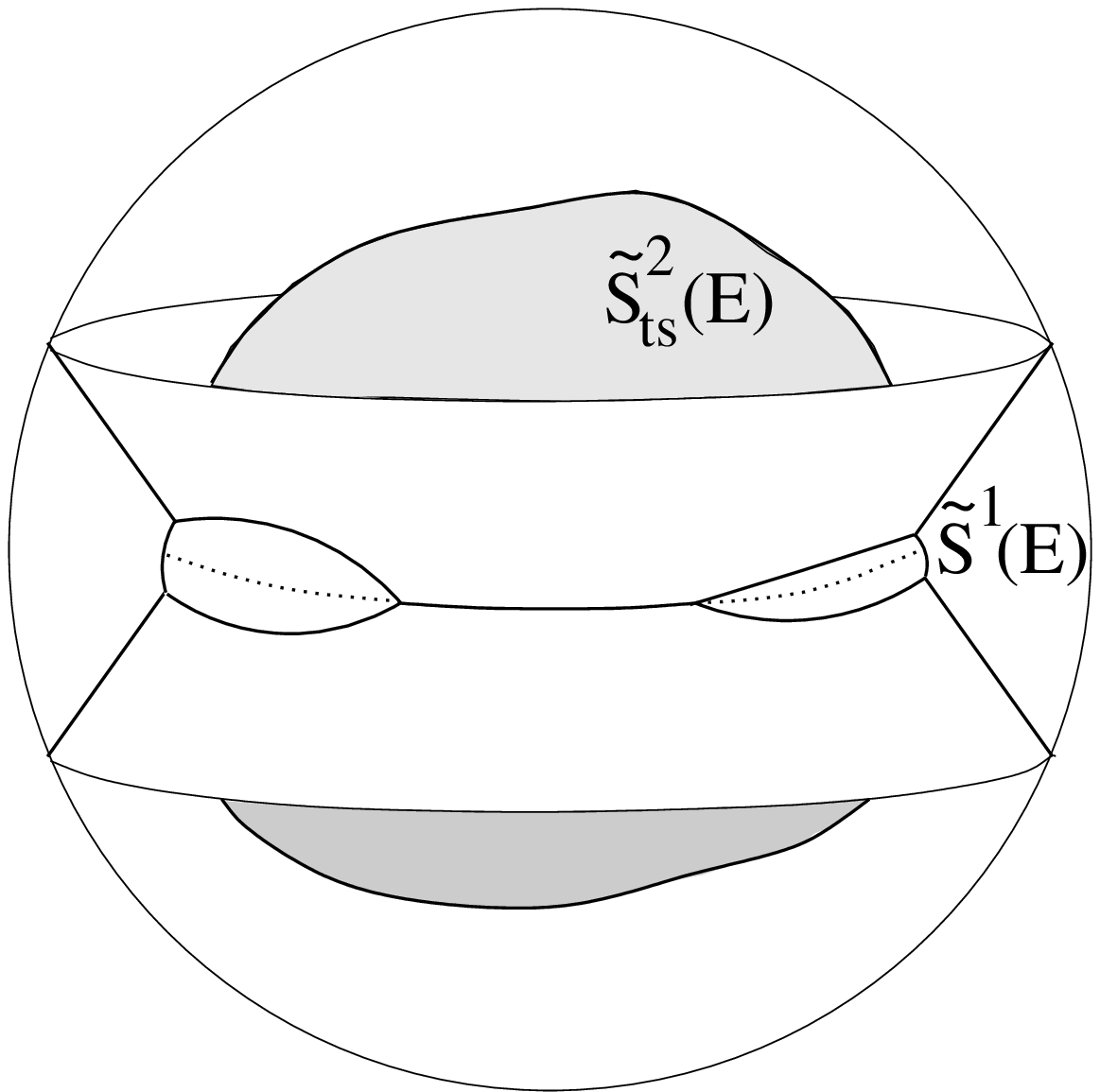} 
\raisebox{3.7cm}{(d)}
\raisebox{0.4cm}{\includegraphics[width=4cm]{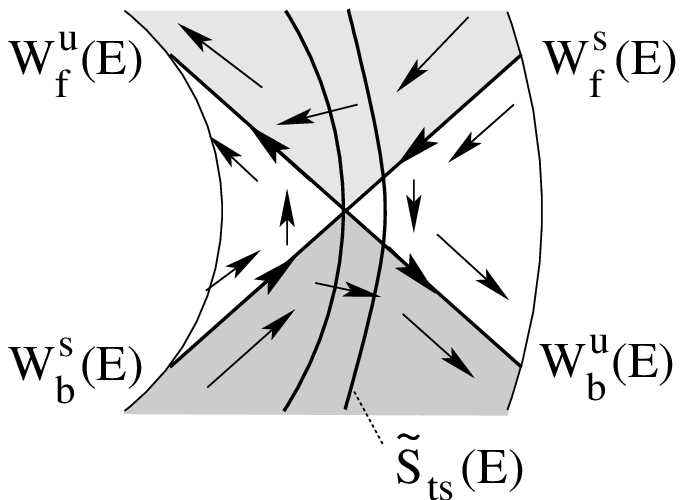}} 
}
\vspace*{0.3cm}
\centerline{
\raisebox{3.2cm}{(e)}
\includegraphics[width=4cm]{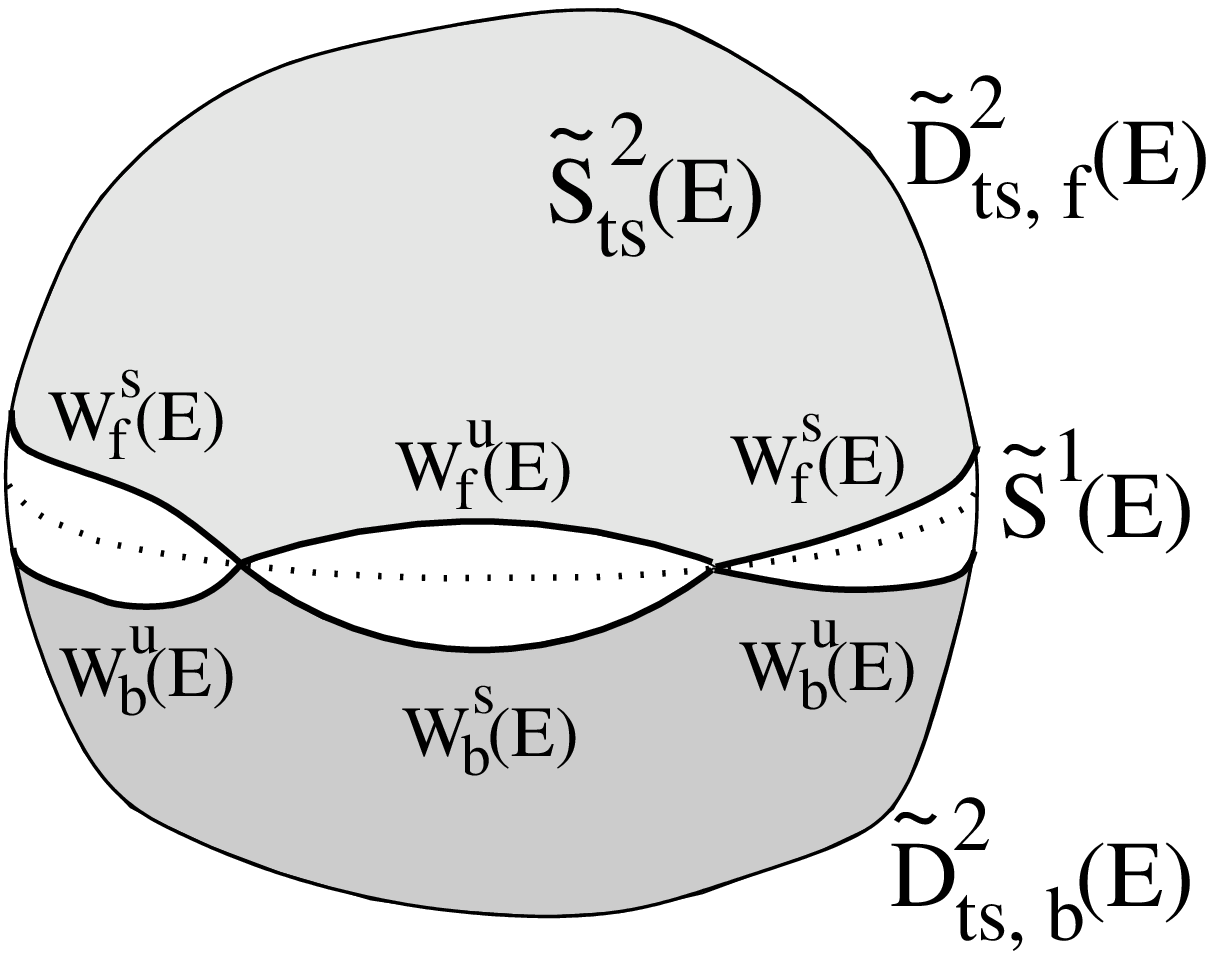}
}
\caption{\label{fig:mcgehee} (a) McGehee representation of the dynamics and the geometry of the manifolds in an energy surface $H=E$ with $E>0$
near a saddle-centre equilibirium point.
The concentric spheres represent $p_1-q_1=c$ (outer sphere), $p_1-q_1=0$ (dividing surface $\tstwoDOF$, middle sphere), and
$p_1-q_1=-c$ (inner sphere, not visible). The equator of the dividing surface is the periodic orbit $\po$.
$W_f^s(E)$, $W_f^u(E)$, $W_b^s(E)$, and $W_b^u(E)$ are the forward and backward branches of its stable and unstable cylinders ($W_b^s(E)$, and $W_b^u(E)$ are not visible).
(b) Section of (a) with a  plane of constant angle about the symmetry axis in (a).
The arrows indicate the Hamiltonian vector field. Note that the vector field also has components out this
the plane. (The energy surface $H=E$ contains no equilibrium points.) Trajectories in the white regions correspond to non-reactive trajectories.
(c) A deformation $\tstwoDOFdef$ of the dividing surface $\tstwoDOF$ and its intersections with
the stable and unstable manifolds of the Lyapunov periodic orbit $\po$. The Hamiltonian vector field is tangent to $\tstwoDOFdef$
along a circle $\podef$ marked by a dotted line (not completely visible).
(d) The same section as in (b) now also showing $\tstwoDOFdef$ in this plane.  
The key point to note here is that  $\podef$ (the intersection of the circle where the vector field is tangent to $\tstwoDOFdef$) moves into the white, non-reactive region.
(e) $\podef$ divides $\tstwoDOFdef$ into two hemispheres
$\tsftwoDOFdef$ and $\tsbtwoDOFdef$. (The extra ``bubbles'' shown in (e), that are not shown in (c),  correspond to the intersection of $\podef$ with $W_b^s(E)$, and $W_b^u(E)$, which are not visible in (c)).}
\end{figure}

Consider at first the case of 2 DOF for which it is 
possible to visualise the energy surface $S^2 \times I$ in 3-dimensional space as a nested set of 2-spheres parametrised along the interval $I$
in radial direction. This is the so-called McGehee representation \cite{McGehee69} which is shown in the first panel of figure~\ref{fig:mcgehee}.
We  consider a  slight {\em generic, iso-energetic} deformation 
of the transition state $\tstwoDOF$ {\em that does not ``preserve'' the saddle sphere $\po$}, i.e., $\po$ is not entirely contained in the deformed transition state, see Figures~\ref{fig:mcgehee}(c)-(e).
The deformation, which we denote by $\tstwoDOFdef$, is described mathematically
as a graph over the transition state $\tstwoDOF$, and therefore $\tstwoDOFdef$ inherits the orientation of $\tstwoDOF$ (which is important because we want to preserve certain aspects of the directionality of the flux, to be described below). 
Because $\tstwoDOFdef$  is chosen to lie in the energy surface, it still separates the energy surface into two disjoint components like $\tstwoDOF$. 
Also like $\tstwoDOF$, the deformation $\tstwoDOFdef$ contains a circle 
$\podef$ of points at which the Hamiltonian vector field is tangent to $\tstwoDOFdef$. This can be  proven analytically, but intuitively it is easy to see from figure~\ref{fig:mcgehee}(d). $\podef$ can be considered as the equator of $\tstwoDOFdef$; 
it divides $\tstwoDOFdef$ into forward and backward hemispheres $\tsftwoDOFdef$ and $\tsbtwoDOFdef$ on which the flux form $\Omega'$ 
is strictly positive and negative, respectively. 
$\podef$ can be considered as the deformation of $\po$ induced by the deformation of $\tstwoDOF$.
It should be realised that in contrast to the case of $\po$, the Hamiltonian vector field is not tangent to the 
deformation $\podef$, i.e. $\podef$ is not invariant under the Hamiltonian vector field ($\podef$ is not a periodic orbit).
It can be shown that if the deformation of $\tstwoDOFdef$ is small enough then all points of $\podef$ have $p_1 q_1\le 0$, 
i.e. they are contained in the complement 
of the two volumes enclosed by the forward and backward reactive spherical cylinders $W_f(E)$ and $W_b(E)$. Again intuitively this can be deduced from  
Figures~\ref{fig:mcgehee}(c)-(e), where the essential point is seen in (d) where  $\tstwoDOFdef$ protrudes into the white, non-reactive region. This is where the trajectories become tangent to  $\tstwoDOFdef$, i.e., at some point on  $\tstwoDOFdef$ trajectories ``bounce off''.

Consider the part of $\tstwoDOFdef$ which is contained in the forward reactive spherical cylinder $W_f(E)$ 
and marked by the
light grey region in figure~\ref{fig:mcgehee}(e). 
The  boundary of the light grey region corresponds to the intersection 
with the forward stable and unstable cylinder branches $W_f^s(E)$ and $W_f^u(E)$.
The boundary is in general not a smooth manifold but it is homeomorphic to a circle $S^1$. It is possible to deform this 
piecewise smooth circle continuously onto the Lyapunov periodic orbit $\po$ without leaving the stable and unstable cylinders. 
Stokes' theorem implies that the difference of the integrals of $\varphi$ along the piecewise smooth circle and the 
Lyapunov periodic orbit is given by the integral of $\Omega'$ over the region on the stable and unstable manifolds
swept out in the deformation process. 
The flux form $\Omega'$ vanishes on the stable and unstable manifolds as its definition involves the interior product with the Hamiltonian vector field 
(``there is no flux through invariant manifolds''). It thus follows that the flux through the 
light grey part of $\tsftwoDOFdef$   in figure~\ref{fig:mcgehee}(d) is equal to
the flux through $\tsftwoDOF$. Therefore since the flux form $\Omega'$ is strictly positive on the complete hemisphere $\tsftwoDOFdef$, 
the forward flux through $\tstwoDOFdef$ is larger than the forward flux through $\tstwoDOF$, and this completes the argument.

All the arguments above immediately carry over to systems with more than 2 DOF by simply adjusting the dimension of the involved manifolds. The essential geometric conditions and relations amongst the manifolds all hold.

The dividing surface of minimal forward flux is not unique, and this is the reason that we chose iso-energetic deformations that did not preserve the NHIM.
For example the equation $p_1=a q_1$ with $a$ (slightly) deviating from 1 again 
defines on the energy surface a $(2n-2)$-sphere which is a dividing surface and 
which has the same forward flux as $\ts$. 
In our consideration we considered a {\em generic} deformation which is  a deformation that changes the equator of $\tsdef$ such that it no longer 
coincides with the NHIM $\nhim$. Non-generic deformation in this sense do not change the flux. Similarly, 
our transition state for 2 DOF coincides with a PODS in general only along its equator, i.e. at the Lyapunov periodic orbit, and this is what matters
for the flux.


\section{Algorithm for computing the flux}

Provided a {\em generic} non-resonance condition between the linear frequencies $\omega_k$ is fulfilled 
the normal form Hamiltonian (\ref{eq:normalform})
assumes the simple form $H({\cal I},J_2,\dots,J_n)$ where $J_k=(p_k^2+q_k^2)/2$, $k=2,\dots,n$, are action variables associated with 
the bath coordinates. Like the saddle integral ${\cal I}$ the actions $J_k$ are constants of the motion. 
Writing the flux form $\Omega'$ in terms of action-angle variables we obtain the result that the forward flux through the transition state
is given by
\[
{\cal F}_E = (2\pi)^{n-1} {\cal V}(E)
\]
where ${\cal V}(E)$ is the volume in the space of the bath actions $(J_2,\dots,J_n)$ enclosed by the contour
$H(0,J_2,\dots,J_n)=E$. In fact the flux can be interpreted as the volume  enclosed by a contour of constant energy $E$ in phase space of a reduced system 
with one dimension less than the complete system. In terms of the normal form coordinates the reduced system is explicitly described by the
invariant subsystem which has $q_1=p_1=0$. The normally hyperbolic invariant manifolds are just the energy surfaces of this reduced system which is 
referred to as {\em activated complex} in the chemistry literature.
The dimensionless quantity ${\cal F}_E/h^{n-1}$, where $h$ is Planck's constant, is 
Weyl's approximation of the integrated density of states of the
reduced system and can be interpreted as the number of ``transition channels''.

In the linear case we have $H=\lambda {\cal I}+ \sum_{k=2}^n \omega_k J_k$ and the energy surface $H=E$ encloses a simplex in 
$(J_2,\dots,J_n)$ whose volume leads to the  well-known result (see e.g. \cite{MacKay1990} for an historical background)
\[
{\cal F}_E = \frac{ E^{n-1}}{(n-1)!} \prod_{k=2}^{n}
\frac{2\pi}{\omega_k}
\]
showing that the flux scales with $E^{n-1}$ for energies close to the saddle energy.
The normal form allows to include the non-linear corrections to this result to any desired order.


\section{Conclusions}

In this letter we have given a geometrical argument  showing that the flux across the dividing surface or {\em transition state} for $n$ DOF systems constructed in \cite{wwju, ujpyw} is a minimum, in the sense that the flux corresponding to forward reactive trajectories (which is equal in magnitude but opposite in sign to the flux corresponding to backward reactive trajectories) is a minimum. 
Hence it is the  optimal dividing surface sought for by 
variational transition theory. 

The key point for the construction of the dividing surface is the existence of a normally hyperbolic invariant manifold (NHIM) which exists on the energy surface near a saddle-center-...-center equilibrium point and which mainly controls the dynamics nearby. The NHIM has the topology of a $(2n-3)$-sphere which can be considered as the equator of the dividing surface which is a $(2n-2)$-sphere. The NHIM divides the dividing surface into two ``hemispheres'' which are open $(2n-2)$-discs. The hemispheres are transverse to the Hamiltonian flow. Hence the dividing surface is everywhere transverse to the Hamiltonian flow except for its equator (the NHIM) which is an invariant manifold. 
The NHIM can be considered as the energy surface of an invariant subsystem (``activated complex'') with one DOF less than the complete system. The flux is the phase space volume enclosed by the energy surface of this reduced system. 

The dividing surface is not unique. 
Any $(2n-2)$-sphere which contains the NHIM as its equator and which is transverse to the Hamiltonian flow except for its equator qualifies for a dividing surface. Nevertheless, all these dividing surfaces lead to the same flux.  
This makes it difficult (if not impossible) to compute a dividing surface from a variational principle of codimension 1 manifolds of the energy surface.
Similarly, the NHIM appears only as an indefinite critical ``point'' of the variational principle of codimension 2 manifolds of the energy surface
of MacKay \cite{MacKay1991} what makes it practically infeasible to compute the NHIM from a variational principle.
The normal form approach of \cite{wwju, ujpyw} is currently the only method to determine these manifolds  and also to compute the flux 
for which we gave an algorithm in this letter.


\section*{Acknowledgments}

This work was  supported by the Office of Naval Research.
H.W. acknowledges support from the Deutsche Forschungsgemeinschaft (Wa 1590/1-1).



\newpage


\begin{thebibliography}{0}

\bibitem{wwju}
Wiggins S, Wiesenfeld L, Jaff\'e C and Uzer T 2001
{\it Phys. Rev. Lett.} {\bf 86} 5478

\bibitem{ujpyw}
Uzer T, Jaff\'e C, Palaci\`an J, Yanguas P and Wiggins S 2002
{\it Nonlinearity} {\bf 15} 957

\bibitem{miller}
Miller W H 1998 {\it Faraday Discuss.} {\bf 110} 1

\bibitem{truhlar1}
Truhlar D G, Garrett, B C and Klippenstein, S J. 1996
{\it J. Phys. Chem.} {\bf 100} 12771



\bibitem{JaffeFarellyUzer2000}
Jaff{\'e} C, Farrelly D and Uzer T  2000
{\it Phys. Rev. Lett.} {\bf 84} 610

\bibitem{KomaBerry1999}
Komatsuzaki T and Berry R S 1999
{\it J. Chem. Phys.} {\bf 110} 9160


\bibitem{Jacucci}
Jacucci G, Toller M, DeLorenzi G and Flynn C P 1984
{\it Phys. Rev. Lett.} {\bf 52} 295



\bibitem{Eckhardt1995}
Eckhardt B. 1995 
{\it J. Phys. A: Math. Gen.} {\bf 28} 3469

\bibitem{cosmo}
de Oliveira H P, Ozorio de Almeida A M, $ \rm Dam\widetilde{ia}o$ Soares I and  Tonini E V 2002
{\it Phys. Rev. D.} {\bf 65} 083511



\bibitem{ross}
Ross S D, Lo M W,  Marsden J, Farrelly D. and Uzer T 2002
{\it Phys. Rev. Lett.} {\bf 89}(1) 011101

\bibitem{Keck1967}
Keck J C 1967
{\it J. Chem. Phys.} {\bf 13} 85

\bibitem{truhlar2}
Truhlar D G and Garrett B C 1984
{\it Ann. Rev. Phys. Chem.} {\bf 35} 159

\bibitem{pm}
Pechukas P and McLafferty F J 1973
{\it J. Chem. Phys.} {\bf 58} (4) 1622

\bibitem{pp1}
Pechukas P and Pollak E 1977
{\it J. Chem. Phys.} {\bf 67}(12) 5976

\bibitem{pp2}
Pechukas P and Pollak E 1978
{\it J. Chem. Phys.} {\bf 69}(3) 1218

\bibitem{pp3}
Pechukas P and Pollak E 1979
{\it J. Chem. Phys.} {\bf 71}(5) 2062

\bibitem{MacKay1990}
MacKay R S 1990
{\it Phys. Lett. A} {\bf 145} 425
  
\bibitem{MacKay1991}
MacKay R S 1991
{\it Nonlinearity} {\bf 4} 155

\bibitem{McGehee69}
McGehee R P 1969
Ph. D. thesis. University of Wisconsin
 

		
\end{thebibliography}
\end{document}